\documentclass[aps,pra,reprint,superscriptaddress]{revtex4-2}

\usepackage{graphicx}
\usepackage{blindtext}
\usepackage{color}
\usepackage{braket}
\usepackage{subfig}
\usepackage{float}

\usepackage{amsmath,amsfonts,amssymb}
\usepackage{xcolor,graphicx}
\usepackage{tikz}
\usepackage[pdftex]{hyperref} 
\usepackage{braket}
\usepackage{dcolumn}
\usepackage{bm}
\hypersetup{
	colorlinks = true,
	linkcolor = blue,
	anchorcolor = blue,
	citecolor = red,
	filecolor = blue,
	urlcolor = blue}

\begin{document}

\title{Discriminando superposiciones de estados coherentes  mediante  formas de línea}

\author{L. Hernández-Sánchez}%
\email[e-mail: ]{leonardi1469@gmail.com}
\affiliation{Instituto Nacional de Astrofísica Óptica y Electrónica, Calle Luis Enrique Erro No. 1\\ Santa María Tonantzintla, Puebla, 72840, Mexico}
\author{I. Ramos-Prieto}%
\affiliation{Instituto Nacional de Astrofísica Óptica y Electrónica, Calle Luis Enrique Erro No. 1\\ Santa María Tonantzintla, Puebla, 72840, Mexico}
\author{F. Soto-Eguibar}%
\affiliation{Instituto Nacional de Astrofísica Óptica y Electrónica, Calle Luis Enrique Erro No. 1\\ Santa María Tonantzintla, Puebla, 72840, Mexico}
\author{H.M. Moya-Cessa}%
\affiliation{Instituto Nacional de Astrofísica Óptica y Electrónica, Calle Luis Enrique Erro No. 1\\ Santa María Tonantzintla, Puebla, 72840, Mexico}
\date{\today}
\begin{abstract}
    Este artículo investiga el efecto de niveles cercanos no resonantes en las líneas espectrales de los átomos que interactúan con un campo electromagnético. Específicamente, examinamos el efecto AC Stark que ocurre cuando la frecuencia del campo coincide con la frecuencia de transición entre dos niveles más bajos y el campo tiene un número promedio pequeño de fotones ($|\alpha|^2 <4$). Nuestra investigación demuestra que los cambios en la forma de la línea espectral se pueden utilizar para distinguir entre los estados de gato de Schrödinger con fases opuestas en $\pi$, a saber, los estados $\ket{\alpha} + \ket{-\alpha}$ y $\ket{\alpha} - \ket{-\alpha}$
\end{abstract}
\maketitle
\section{Introducción}
Durante las últimas décadas ha habido un considerable interés en las propiedades de los estados superpuestos de la luz, especialmente en la superposición de dos estados coherentes \cite{Dodonov_1974,Mandel_1986, Buzek_1992b, Moya_1995a,Yurke_1986,Wodkiewicz_1987,Vidiella_1992,Janszky_1995,Gerry_1997,Recamier_2000,Recamier_2003,Segundo_2003,Muhammad_2009}, debido a que exhiben propiedades muy distintas a las de los estados que los componen. Estas propiedades incluyen la compresión en cuadratura~\cite{Buzek_1992b}, el tipo de estadística (sub-poissoniana o super-poissoniana) \cite{Mandel_1986,Gerry_1997}, y la capacidad para caracterizar estados no clásicos de la luz~\cite{Dodonov_1974,Mandel_1986, Buzek_1992b, Moya_1995a}, entre muchas otras~\cite{Yurke_1986,Wodkiewicz_1987,Vidiella_1992,Janszky_1995,Gerry_1997,Recamier_2000,Recamier_2003,Segundo_2003,Muhammad_2009}.\\
En el modelo de Jaynes-Cummings~\cite{JC_1963}, el problema de la interacción entre un solo modo del campo electromagnético, preparado en una superposición de estados coherentes, y un átomo de dos niveles ya ha sido abordado en estudios previos~\cite{Buzek_1992b, Vidiella_1992}. Estos estudios han demostrado que el tiempo de colapso y resurgimiento de la inversión atómica se reduce a la mitad en comparación con un estado coherente~\cite{Eberly_1980}. Cabe mencionar que se han desarrollado generalizaciones del modelo de Jaynes-Cummings para abordar casos específicos, como la interacción con un campo inicialmente preparado en un estado coherente comprimido~\cite{Dodonov_1990, Moya_1992}, la interacción con un medio no lineal de tipo Kerr~\cite{Santos_2012}, las interacciones con dos excitaciones~\cite{Ramos_2014}, las interacciones de tipo optomecánico~\cite{Medina_2020}, y el acoplamiento entre dos hamiltonianos de Jaynes-Cummings utilizando álgebras de Lie~\cite{Ramos_2020}.\\
Por otro lado, en trabajos experimentales de micromaser~\cite{Meschede_1985}, se han observado asimetrías y cambios en las formas de línea; es decir, se ha observado una variación de la inversión atómica promedio en función de la desintonía, la cual se ha atribuido a los efectos de los niveles cercanos no resonantes~\cite{Meschede_1985}. Sin embargo, para poder visualizar la firma estadística del campo en la dinámica atómica, fenomenológicamente se agrega un término de tipo AC Stark que genera niveles virtuales cerca de la resonancia~\cite{Moya_1991}. Aunque se han ideado diversas estrategias para poder discriminar entre dos o más estados cuánticos en superposición, únicamente en algunos casos se ha podido establecer un mecanismo para diferenciarlos~\cite{Bae_2015, Barnett_2001, Barnett_2009, Bergou_2007, Chefles_2000}. La posibilidad de discernir entre una superposición de estados coherentes con una distribución de fotones par o impar, mediante las formas de línea, es la motivación de este artículo.\\
En este trabajo, abordamos el estudio del modelo de Jaynes-Cummings con el término de AC Stark que considera los niveles cercanos no resonantes. En la sección~\ref{Sec1} resolvemos la ecuación de Schrödinger, y consideramos los casos particulares de cuando inicialmente el átomo está en su primer estado excitado, y cuando el átomo originalmente está en el estado base. Tomando como condición inicial del campo un estado coherente, en la sección~\ref{Coherentes}, analizamos los efectos de los niveles cercanos no resonantes en la inversión atómica y mostramos cómo las formas de línea se ensanchan a medida que el número promedio de fotones aumenta. En la sección~\ref{Sec2}, extendemos el análisis a una superposición de estados coherentes par e impar con las mismas condiciones atómicas que en la sección~\ref{Coherentes}. Mostramos ahora que, al tomar en cuenta los niveles cercanos no resonantes y para un número promedio de fotones suficientemente pequeño, las formas de línea permiten discernir entre un estado gato de Schrödinger par e impar. Finalmente en la sección~\ref{Conclusión} escribimos nuestras conclusiones.
\section{Modelo de Jaynes-Cummings y estados virtuales no resonantes}\label{Sec1}
\begin{figure}
    \centering
    \includegraphics[width = \linewidth]{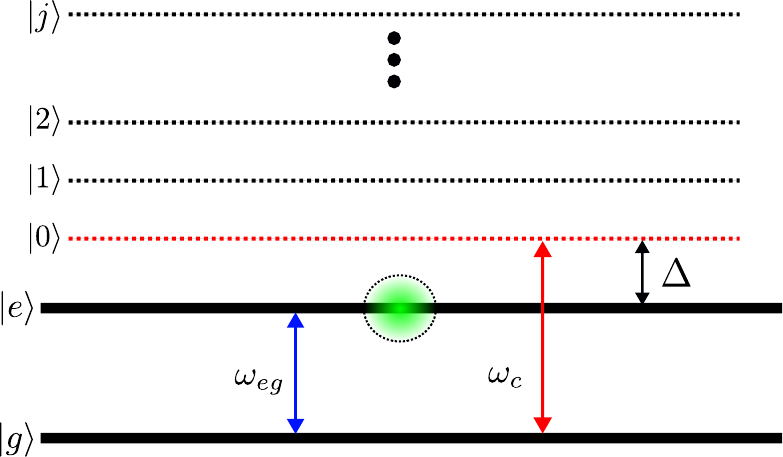}
    \caption{Esquema de niveles que indican el par de estados atómicos excitados casi resonantes con frecuencia de transición $\omega_{eg}$, la frecuencia de campo $\omega_{c}$, y un conjunto de niveles no resonantes que participan solo virtualmente en la excitación y son responsables de los cambios de AC Stark a la frecuencia de transición $\omega_{eg}$.}
    \label{fig1}
\end{figure}

Consideremos un átomo con un estado base $\ket{g}$, un estado excitado $\ket{e}$ y estados superiores denotados por $\ket{j}$, con $j = 0,1,2,...,\infty$. El átomo interactúa con un campo de un solo modo, como se muestra en la figura~\ref{fig1}. Suponemos que el campo está aproximadamente en sintonía con la frecuencia de transición entre los niveles $\ket{g}$ y $\ket{e}$ del átomo, pero fuera de sintonía con los niveles cercanos $\ket{j}$ (efecto AC Stark).
El Hamiltoniano que describe este sistema se expresa como~\cite{Meschede_1985,Moya_1991,Moya_1995b,Villanueva_2020}
\begin{equation} \label{eq:1}
    \hat{H}  = \frac{\omega_{eg}}{2} \hat{\sigma}_{z} +  \omega_c\hat{a}^{\dagger} \hat{a}+ \chi \hat{a}^{\dagger} \hat{a} \hat{\sigma}_{z}+ g \left( \hat{\sigma}_{+}\hat{a} + \hat{\sigma}_{-} \hat{a}^{\dagger} \right),
\end{equation}
donde $g$ es la constante de acoplamiento entre el sistema de dos niveles y el campo (en la aproximación dipolar), mientras que $\chi$ es el parámetro que cuantifica la intensidad de la interacción en el efecto AC Stark, debido a la presencia de niveles virtuales cercanos no resonantes. Se utilizan los operadores de creación y aniquilación, $\hat{a}^{\dagger}$ y $\hat{a}$, que satisfacen la relación de conmutación $\left[\hat{a}, \hat{a}^{\dagger} \right] =1$. Además, para describir la parte atómica del sistema, se utilizan los operadores $\hat{\sigma}_{+} = \ket{e}\bra{g}$, $\hat{\sigma}_{-} = \ket{g}\bra{e}$ y $\hat{\sigma}_{z} = \ket{e}\bra{e} - \ket{g}\bra{g}$, los cuales cumplen las relaciones de conmutación $[\hat{\sigma}_{+}, \hat{\sigma}_{-}] = \hat{\sigma}_z$ y $[\hat{\sigma}_{z}, \hat{\sigma}_{\pm}] = \pm 2 \hat{\sigma}_{\pm}$.\\
Para resolver la ecuación de Schrödinger de este sistema, realizamos la transformación unitaria dependiente del tiempo $\hat{\mathcal{R}}=\exp\left[\mathrm{i} \omega_c t (\hat{n}+ \hat{\sigma}_z /2)\right]$, que nos conduce a la representación de interacción, cuyo Hamiltoniano es
\begin{equation}
    \begin{split}
        \hat{\mathcal{H}} &= \hat{\mathcal{R}}\hat{H} \hat{\mathcal{R}}^{\dagger}-\mathrm{i}\hat{\mathcal{R}}\partial_t\hat{\mathcal{R}}^\dagger,\\
        &=\left( \frac{\Delta}{2} + \chi \hat{n} \right) \hat{\sigma}_z + g \left( \hat{\sigma}_{+}\hat{a} + \hat{\sigma}_{-} \hat{a}^{\dagger} \right),
    \end{split}
\end{equation}
donde $\Delta = \omega_{eg} - \omega_c$ es la desintonía entre la frecuencia de campo y la frecuencia de transición atómica.\\
Para resolver la ecuación de Schrödinger en el esquema de interacción, utilizamos el método tradicional~\cite{Klimov_2009,Amaro_2015}, que consiste en proponer un desarrollo del vector de estado átomo-campo al tiempo $t$ como una combinación lineal o una superposición de estados de Fock $\left\{ \ket{n} \right\}$. Esta superposición se puede escribir como
\begin{equation}\label{0030}
    \ket{ \Psi (t) }=\sum_{n=0}^{\infty}\left[C_n (t) \ket{n} \ket{e}+D_n(t) \ket{n+1} \ket{g} \right],
\end{equation}
y el problema se reduce a resolver el siguiente sistema de ecuaciones diferenciales ordinarias acopladas
\begin{align}
    \mathrm{i}\frac{d}{dt}
    \begin{bmatrix}
        C_n(t) \\D_n(t)\\
    \end{bmatrix}
    = &
    \begin{bmatrix}
        \chi n +\frac{\Delta}{2} & g\sqrt{n+1}                 \\
        g\sqrt{n+1}              & -\chi(n+1)-\frac{\Delta}{2} \\
    \end{bmatrix}
    \begin{bmatrix}
        C_n(t) \\D_n(t)\\
    \end{bmatrix}.
    \nonumber \\ &
    n=0,1,2,\dots.
\end{align}
La solución general de estas ecuaciones diferenciales es
\begin{align}\label{0050}
    \begin{bmatrix}
        C_{n}(t) \\D_n(t)
    \end{bmatrix}
    = & \exp\left(\mathrm{i}  \frac{\chi t}{2}\right)
    \begin{bmatrix}
        M_{11}(t) & M_{12}(t) \\
        M_{21}(t) & M_{22}(t) \\
    \end{bmatrix}
    \begin{bmatrix}
        C_{n}(0) \\
        D_{n}(0)
    \end{bmatrix},
    \nonumber                                         \\ &
    n=0,1,2,\dots,
\end{align}
donde
\begin{equation}
    \begin{split}\label{C_nD_n}
        M_{11}(t) &=\cos\left(\frac{\beta_n t}{2} \right) -\mathrm{i}\frac{\Delta + \chi(2n+1)}{\beta_n}   \sin\left(\frac{\beta_n t}{2} \right),\\
        M_{12}(t) &=-\mathrm{i}\frac{ 2g \sqrt{n+1}}{\beta_n} \sin\left(\frac{\beta_n t}{2}\right),
        \\  M_{22}(t)&= M_{11}^*(t), \quad M_{21}(t)=M_{12}(t), \quad   n=0,1,2,\dots.
    \end{split}
\end{equation}
Las cantidades $|C_n(0)|^2$ y $|D_n(0)|^2$ determinan la distribución inicial de fotones del campo en el estado excitado y estado base del átomo, respectivamente. Mientras que $\beta_n$,
\begin{equation} \label{Rabi_G}
    \beta_n= \sqrt{\left[ \Delta + \chi (2n+1)  \right]^{2}  + 4g^2 (n+1) },
\end{equation}
es la frecuencia generalizada de Rabi debida a los cambios de AC-Stark,  que son las variaciones en la energía de un átomo debido a la presencia de un campo eléctrico no resonante.\\
Una vez dada la condición inicial átomo-campo $\ket{\Psi(0)}$, es posible obtener la evolución temporal de cualquier observable del sistema. En este caso, nos enfocamos en la inversión atómica, $W(t) = \braket{\Psi(t)|\hat{\sigma}_z|\Psi(t)}$, la cual determina los cambios atómicos de población y contiene la firma estadística del campo. Así, la probabilidad de que el átomo esté en su estado excitado menos la probabilidad de que esté en el estado base se determina de forma general por la siguiente expresión
\begin{equation}\label{InversionW}
    W(t) = \sum_{n=0}^{\infty}  \left(\vert C_n (t) \vert^{2} - \vert D_n (t) \vert^{2} \right).
\end{equation}
Utilizando la solución dada en \eqref{0050}, es sencillo escribir
\begin{widetext}
\begin{align}
    W(t)= & \sum_{n=0}^{\infty}\left[ \left(\left|M_{11}(t)\right|^2-\left|M_{12}(t)\right|^2\right)
    \left(\left|C_n(0)\right|^2-\left| D_n(0)\right|^2\right)
    \right. \nonumber                                                                                \\   &  \left.
    +2 M_{12}(t)
    \left(C_n(0)^* D_n(0)  M_{11}(t)^* -C_n(0) D_n(0)^* M_{11}(t) \right)\right] ;
\end{align}
y si ahora sustituimos los valores de los coeficientes dados en \eqref{C_nD_n}, obtenemos
\begin{align}\label{0100}
    W(t)= & \sum_{n=0}^{\infty}\frac{1}{\beta_n^2}\left\lbrace \left[ \Delta +\left( 2 n+1\right)\chi\right]^2+4g^2 (n+1) \cos \left(\beta_n t \right)\right\rbrace
    \left(\left|C_n(0)\right|^2-\left| D_n(0)\right|^2\right)
    \nonumber                                                                                                                                                       \\ &
    -\sum_{n=0}^{\infty}\frac{4g \sqrt{n+1}}{\beta_n^2}\left[ \Delta +\left( 2 n+1\right)\chi\right] \left(\cos \left(\beta_n t\right)-1\right)
    C_n(0) D_n(0).
\end{align}
\end{widetext}
Si suponemos que el átomo se encuentra inicialmente en su estado excitado, es decir, $\ket{ \Psi (0) }=\sum_{n=0}^{\infty} C_n (0) \ket{n} \ket{e}$, ($D_n(0)=0$ para  $n=0,1,2,\dots$) podemos obtener la inversión atómica de la siguiente manera
\begin{equation}\label{W_e}
    \begin{split}
        W_\textrm{e} (t) = \sum_{n=0}^{\infty}\frac{P_n}{\beta_n^2}
        \bigg\{&\left[ \Delta+(2n+1)\chi\right]^2\\ &+ 4g^2 (n+1)\cos (\beta_{n} t)  \bigg\},
    \end{split}
\end{equation}
donde hacemos la identificación $\left|C_n(0)\right|^2=P_n$ para $n=0,1,2,\dots$, siendo $P_n$ la distribución de probabilidad de fotones.\\
Si suponemos ahora que el átomo está inicialmente en su estado base, o sea, $\ket{ \Psi (0) }=\sum_{n=0}^{\infty} D_n (0) \ket{n+1} \ket{g}$ ($C_n(0)=0$ para  $n=0,1,2,\dots$), la inversión atómica es
\begin{equation}\label{W_g}
    \begin{split}
        W_\textrm{b} (t)=  - \sum_{n=0}^{\infty}\frac{P_{n+1}}{\beta_n^2}
        \bigg\{&\left[ \Delta+(2n+1)\chi\right]^2\\ &+ 4g^2 (n+1)\cos (\beta_{n} t)  \bigg\},
    \end{split}
\end{equation}
donde ahora debemos identificar $|D_n(0)|^2=P_{n+1}$ para $n=0,1,2,\dots$. Esta última identificación tiene sentido físico: si analizamos la expresión \eqref{0030}, que nos da la función de onda del sistema completo, nos damos cuenta que desde un principio hemos supuesto que hay un cuanto de energía, y por lo tanto, si el átomo está en el estado base, la probabilidad de que en el campo haya cero fotones es nula.\\
Una manera de poder analizar las posibles variaciones de las probabilidades de transición entre el nivel base y el primer nivel excitado en función de la desintonía, es usando las formas de línea, que no dependen de la duración del tiempo de interacción $t$. Nos centramos en la inversión atómica promedio $\overline{W}(\Delta)$\cite{Moya_1991}
\begin{equation} \label{W_Delta}
    \overline{W}(\Delta) = \lim_{T \to{\infty}} \frac{1}{T} \int_{0}^{T} W(t) \, dt.
\end{equation}
Dado que
\begin{equation}
    \lim_{T \to{\infty}} \frac{1}{T} \int_{0}^{T} \cos (\beta_{n} t) \, dt=0,
\end{equation}
y usando \eqref{0100}, tenemos que
\begin{align}
    \overline{W}(\Delta) & = \sum_{n=0}^{\infty}\left[ \frac{\Delta +\left( 2 n+1\right)\chi}{\beta_n}\right]^2
    \left(\left|C_n(0)\right|^2-\left| D_n(0)\right|^2\right)
    \nonumber                                                                                                   \\ &
    +\sum_{n=0}^{\infty}4g \sqrt{n+1} \, \left[ \frac{\Delta +\left( 2 n+1\right)\chi}{\beta_n^2}\right]C_n(0) D_n(0).
\end{align}
En el caso en que el átomo se encuentra inicialmente en el estado excitado, obtenemos
\begin{equation}\label{W_Delta_e}
    \overline{W}_\textrm{e}(\Delta) =\sum_{n=0}^{\infty} P_{n}  \left[\frac{\Delta +(2n+1)\chi}{\beta_n}\right]^2,
\end{equation}
mientras que cuando el átomo se encuentra inicialmente en el estado base utilizamos la ecuación \eqref{W_g}, y llegamos a la expresión
\begin{equation}\label{W_Delta_g}
    \overline{W}_\textrm{b}(\Delta) =-\sum_{n=0}^{\infty} P_{n+1}  \left[\frac{\Delta +(2n+1)\chi}{\beta_n}\right]^2.
\end{equation}

\section{Estados coherentes}\label{Coherentes}
\begin{figure}
    \centering
    \subfloat[Distribución de los fotones en un estado coherente con $\alpha=4$.\label{fig2a}]
    {\includegraphics[width = \linewidth]{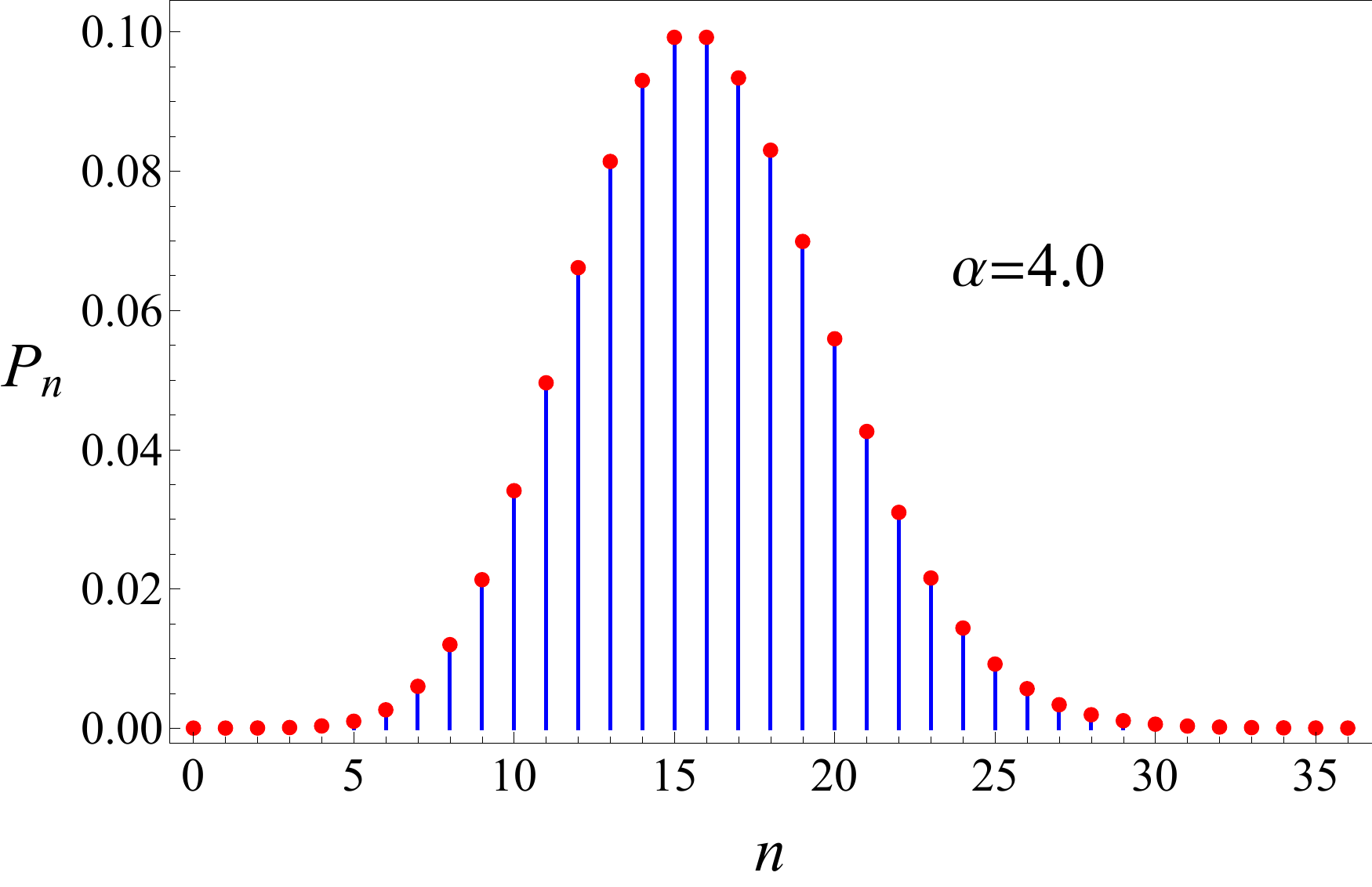}}\\
    \subfloat[Inversión atómica cuando $\chi=0.0$\label{fig2b}]
    {\includegraphics[width = \linewidth]{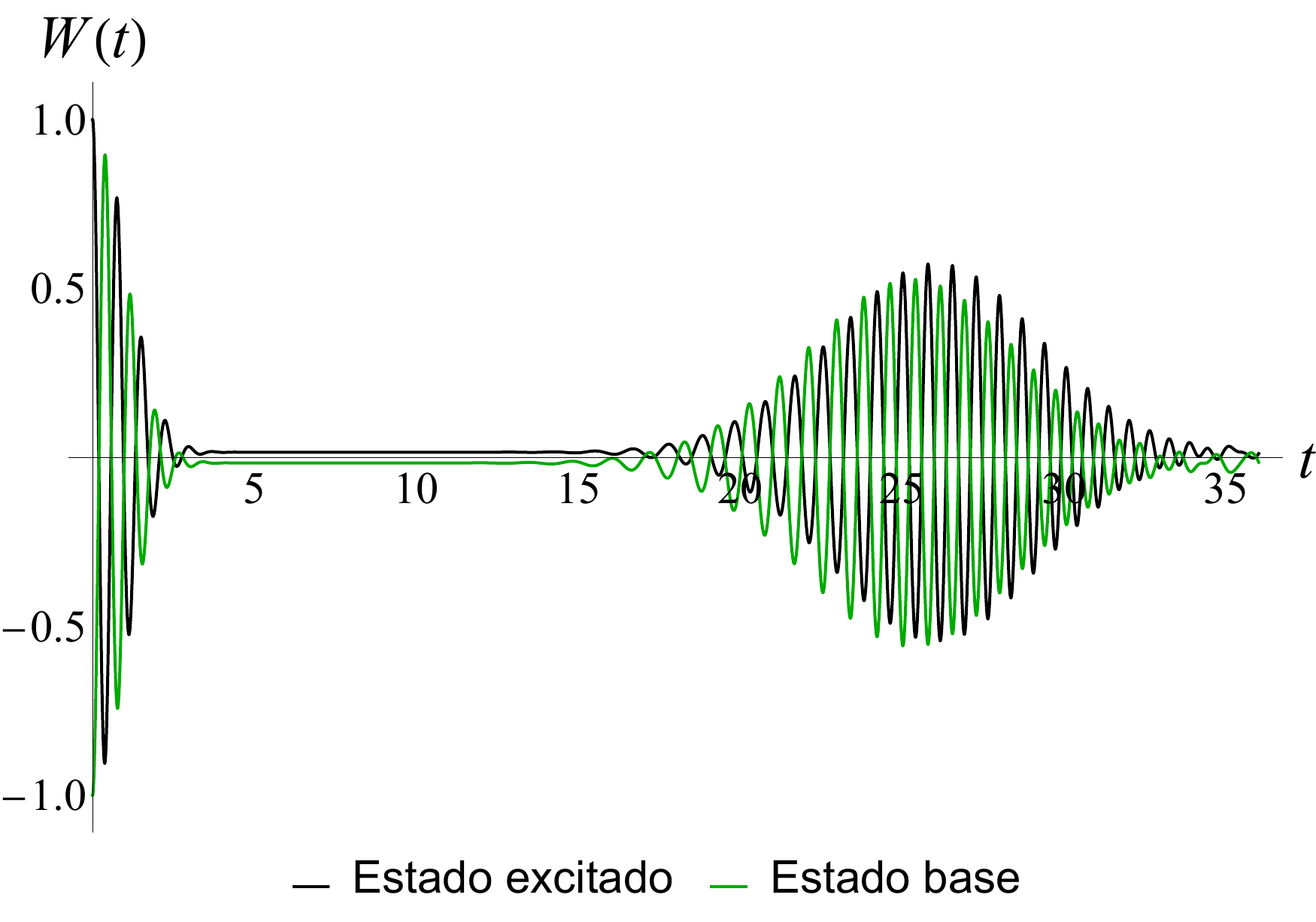}}\\
    \subfloat[Inversión atómica cuando $\chi=0.5$\label{fig2c}]
    {\includegraphics[width = \linewidth]{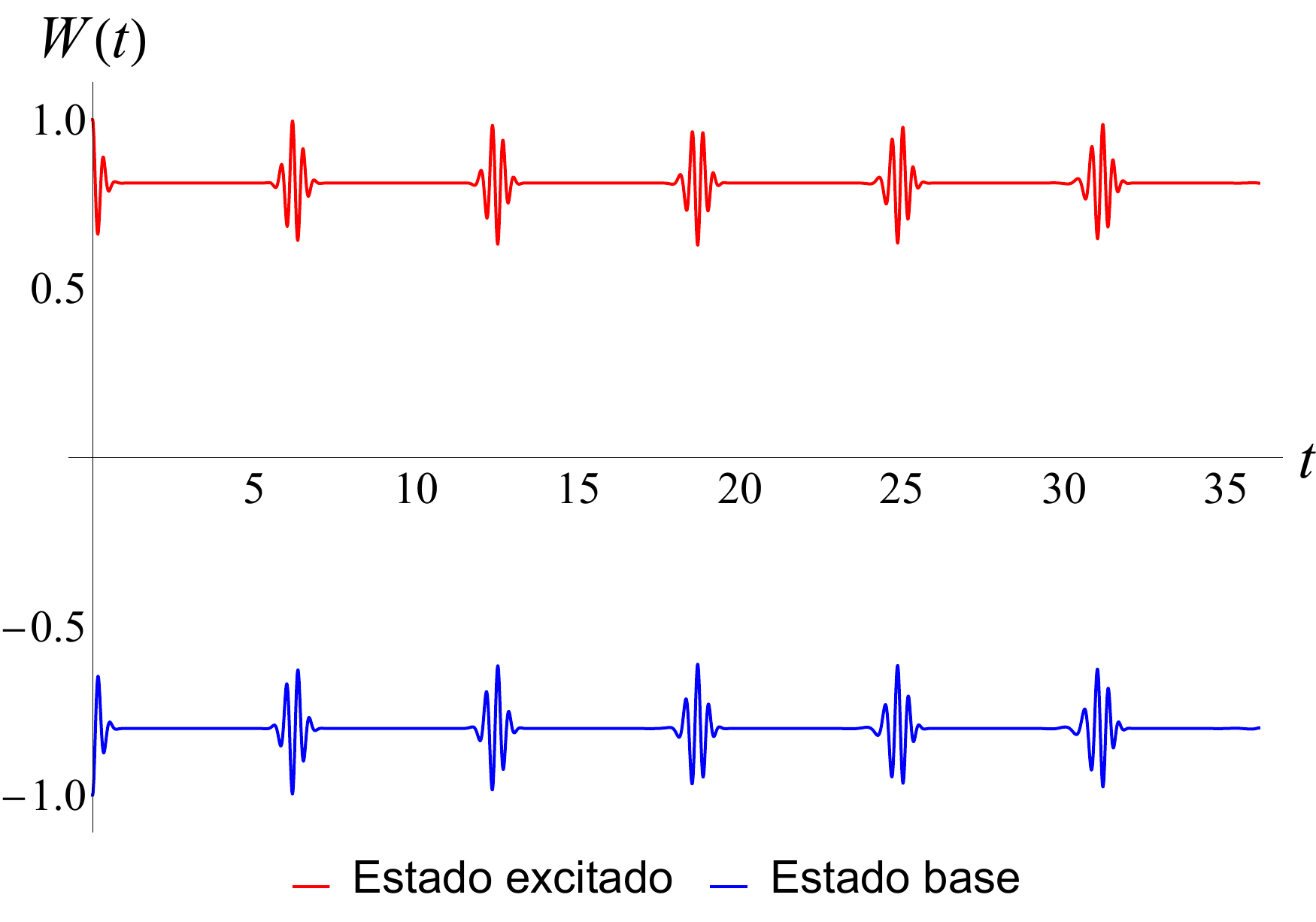}}
    \caption{Gráficas de la inversión atómica cuando inicialmente el campo está en un estado coherente con $\alpha=4$. Los valores de los otros parámetros son $\Delta=1$, y $g=1$.}
    \label{fig2}
\end{figure}

\begin{figure}
    \centering
    \subfloat
    {\includegraphics[width = \linewidth]{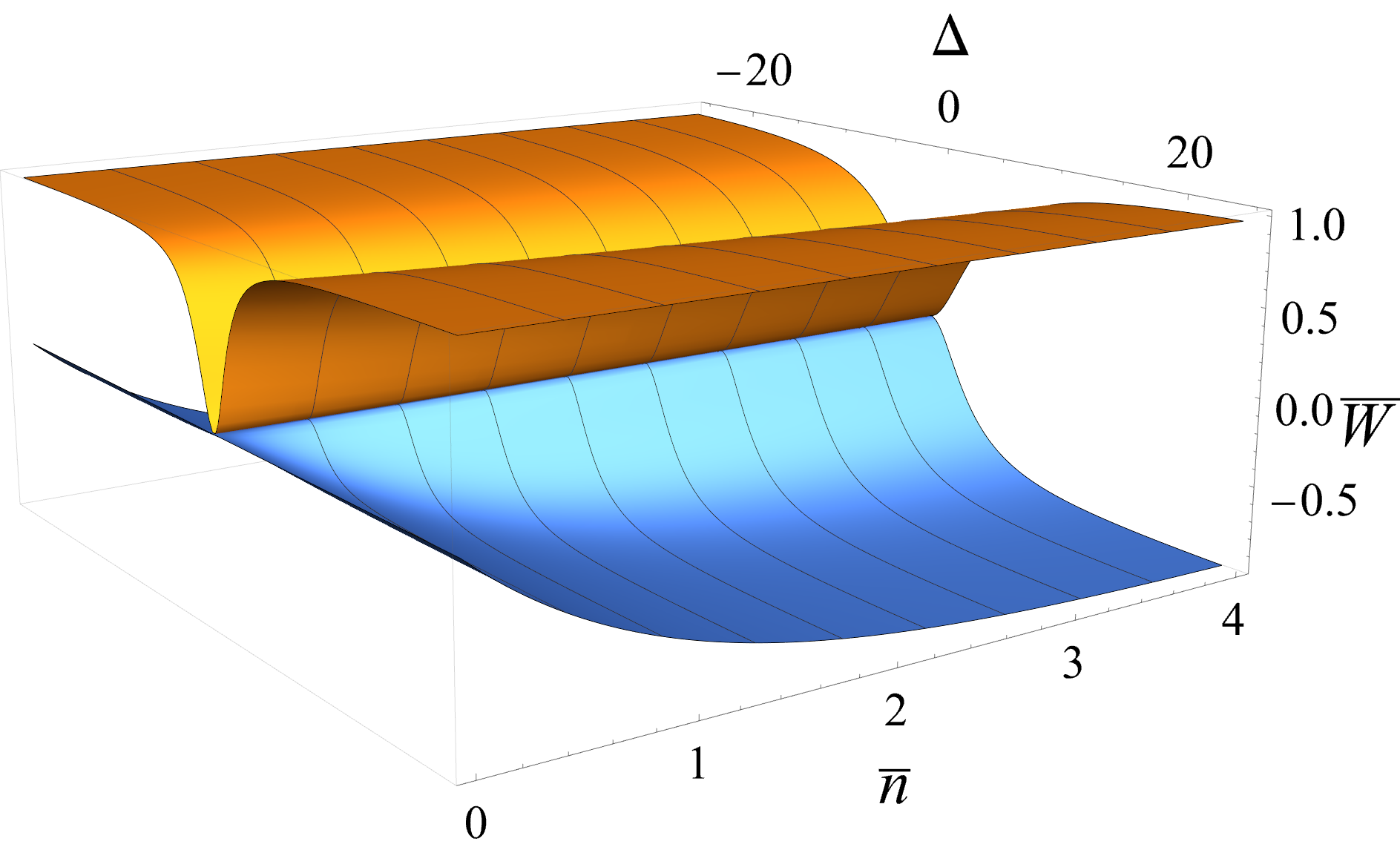}}
    \quad
    \subfloat
    {\includegraphics[width = \linewidth]{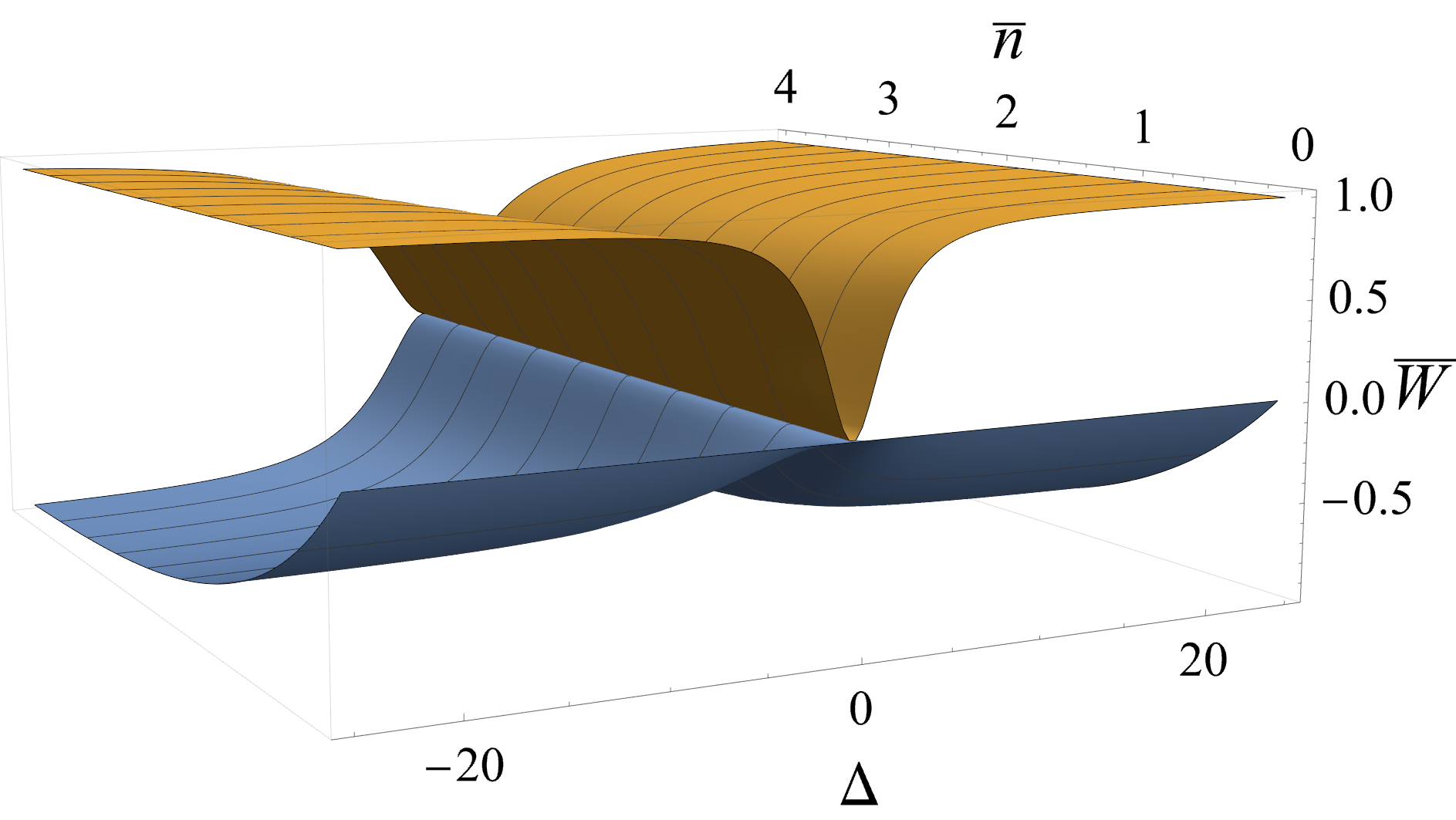}}
    \caption{Inversión atómica promedio $\overline{W}(\Delta)$ para la condición inicial del átomo en el estado excitado (superficie naranja) y para cuando inicialmente el átomo está en su estado base (superficie azul). Las gráficas muestran la inversión atómica promedio como función de $\Delta$ y de $\overline{n}$. Los valores de los parámetros son $\chi=0$ y $g=1.0$.}
    \label{WM3d.EstCoh.chi=0}
\end{figure}

Consideramos ahora que el campo  está inicialmente en un estado coherente $\ket{\alpha}$; por lo tanto, la distribución de fotones es
\begin{equation} \label{P_n_coherente}
    P_{n} = e^{-\bar{n}} \frac{\bar{n}^{n}}{n!}, \qquad n=0,1,2,\dots,
\end{equation}
donde $\bar{n}= |\alpha|^2 $ es el número promedio de fotones. En la figura~\ref{fig2} mostramos la inversión atómica $W(t)$ cuando $\alpha=4$, y para el átomo consideramos dos casos, cuando inicialmente está en su estado excitado, y cuando inicialmente esta en su estado base. Primeramente, en la gráfica \ref{fig2a}, mostramos la distribución de fotones. En la gráfica \ref{fig2b} mostramos la inversión atómica $W(t)$ cuando $\chi = 0$; observamos el colapso y resurgimiento convencionales del modelo de Jaynes-Cummings \cite{Eberly_1980}. Sin embargo, cuando se tienen en cuenta los niveles cercanos no resonantes con $\chi = 0.5$, gráfica \ref{fig2c}, la inversión atómica se acerca en promedio a su valor inicial debido a que los niveles fuera de resonancia suprimen la eficacia del campo para estimular transiciones fuera de su estado inicial. Además, se puede observar que el tiempo para que aparezca el primer resurgimiento se acorta al aumentar los valores de $\chi$ en ambas condiciones iniciales del átomo \cite{Moya_1991}.
En la figura~\ref{WM3d.EstCoh.chi=0} graficamos la inversión atómica promedio $\overline{W}(\Delta)$ para la condición inicial del átomo en el estado excitado (superficie naranja) y para cuando inicialmente el átomo está en su estado base (superficie azul); las gráficas muestran la inversión atómica promedio como función de $\Delta$, que es la desintonía, y de $\overline{n}$, que es el número promedio de fotones. Observamos que las formas de línea se van ensanchando a medida que aumentamos el número promedio de fotones $\bar{n}=|\alpha|^2$, sin embargo, estas mantienen su forma y simetría respecto al origen. En el caso en que el átomo está inicialmente en su estado base (superficie azul), la presencia de una excitación adicional entre el estado base y el estado excitado, causa que el pico máximo en $\Delta=0$ aumente su altura a medida que aumenta el número promedio de fotones, llegando a su punto máximo alrededor de $\bar{n}\approx 4$.

\section{Estados gato de Schrödinger} \label{Sec2}
\begin{figure}
    \centering
    \subfloat[$\alpha=4, \; \phi=0$]
    {\includegraphics[width = \linewidth]{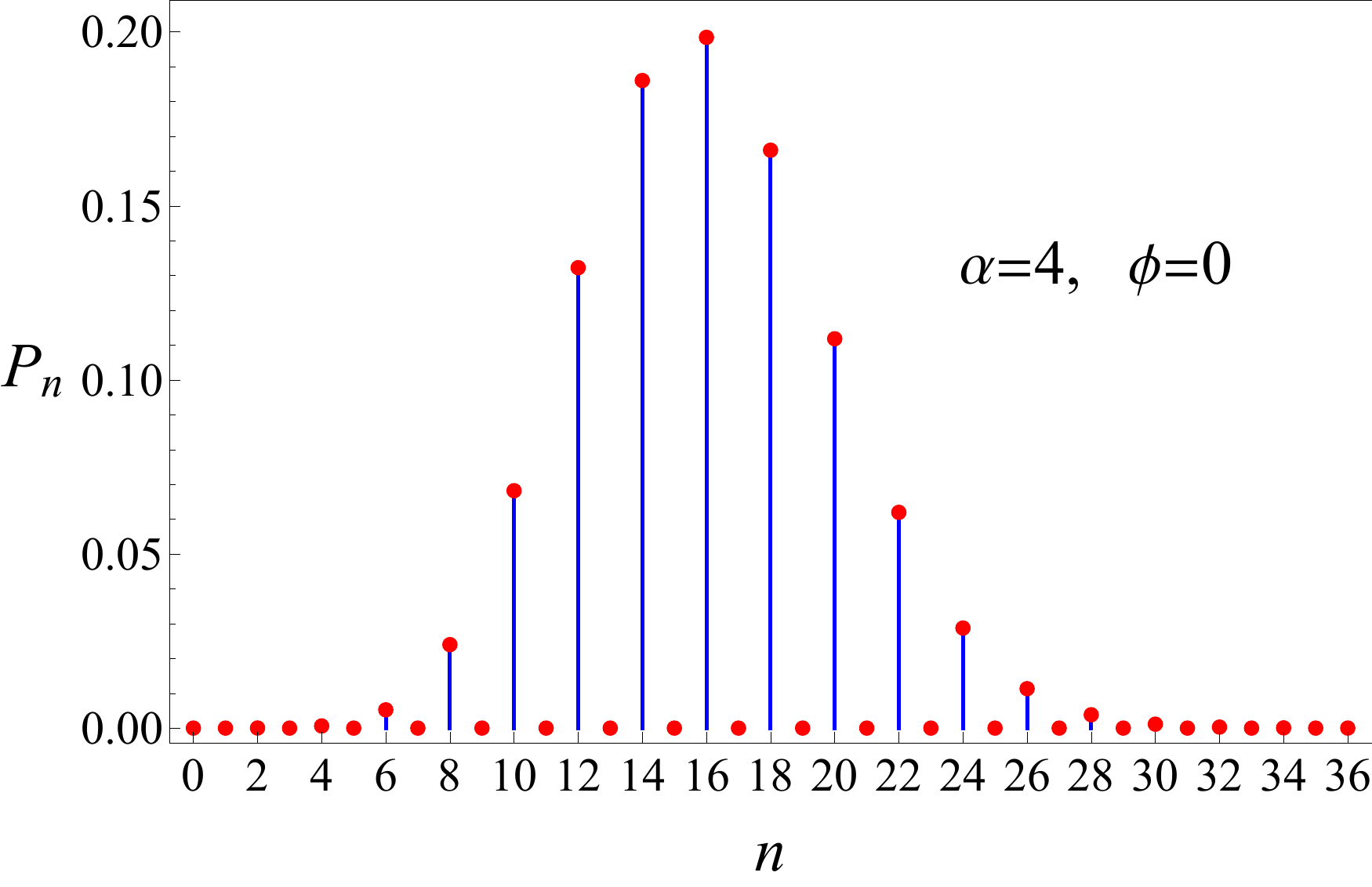}}
    \quad
    \subfloat[$\alpha=4, \; \phi=\pi$]
    {\includegraphics[width = \linewidth]{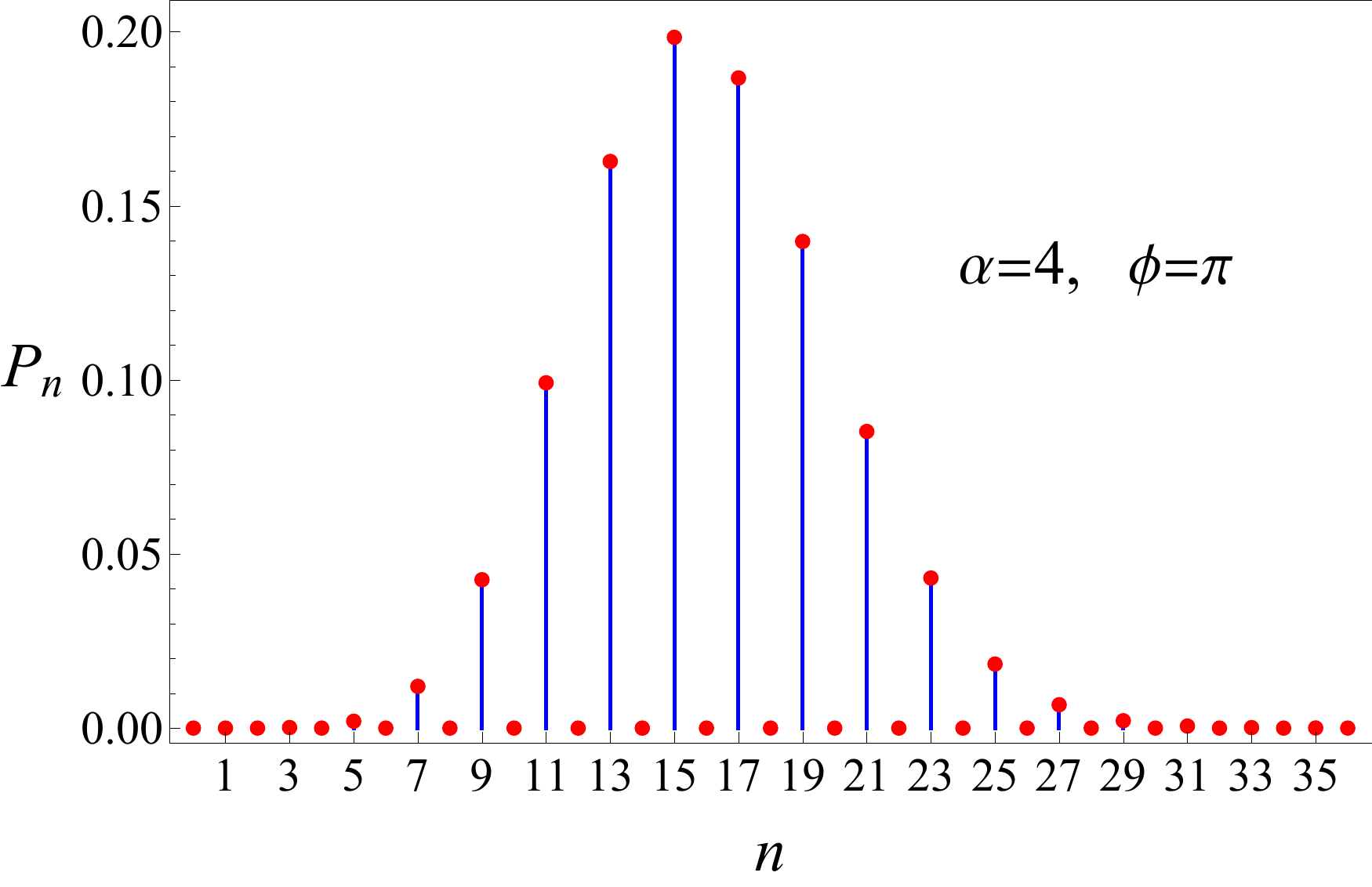}}
    \caption{La distribución de probabilidad de fotones para los estados gato de Schrödinger.}
    \label{fig5}
\end{figure}

\begin{figure}
    \centering
    {\includegraphics[width = \linewidth]{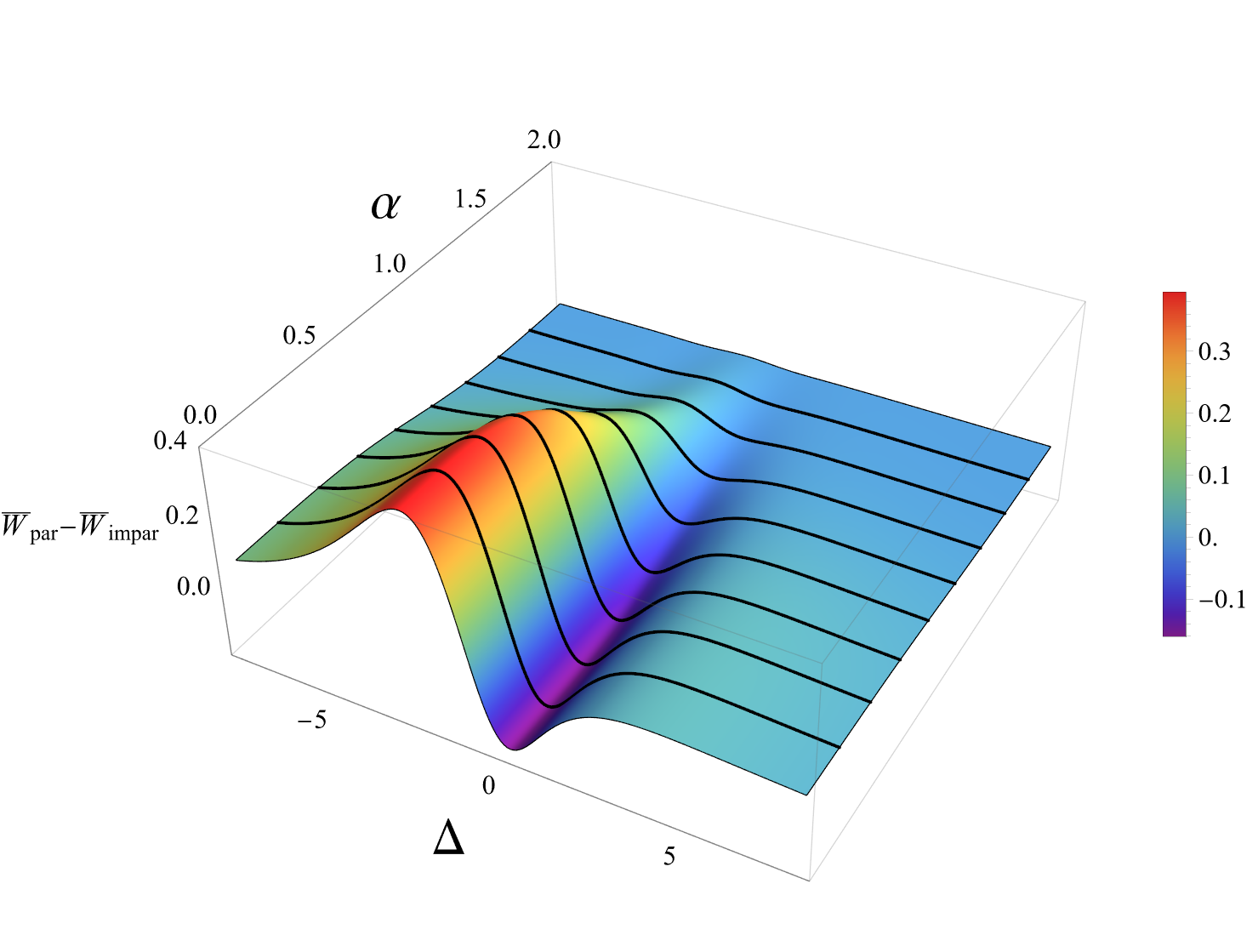}}
    \caption{Diferencia de las formas de línea cuando el estado inicial del átomo es el excitado, y el del campo son los estados gato de Schrödinger par e impar con $\alpha$ variando desde 0 hasta 2. Los valores de los parámetros son $\chi=0.5$ y $g=1.0$.}
    \label{fig6}
\end{figure}

\begin{figure}
    \centering
    {\includegraphics[width = \linewidth]{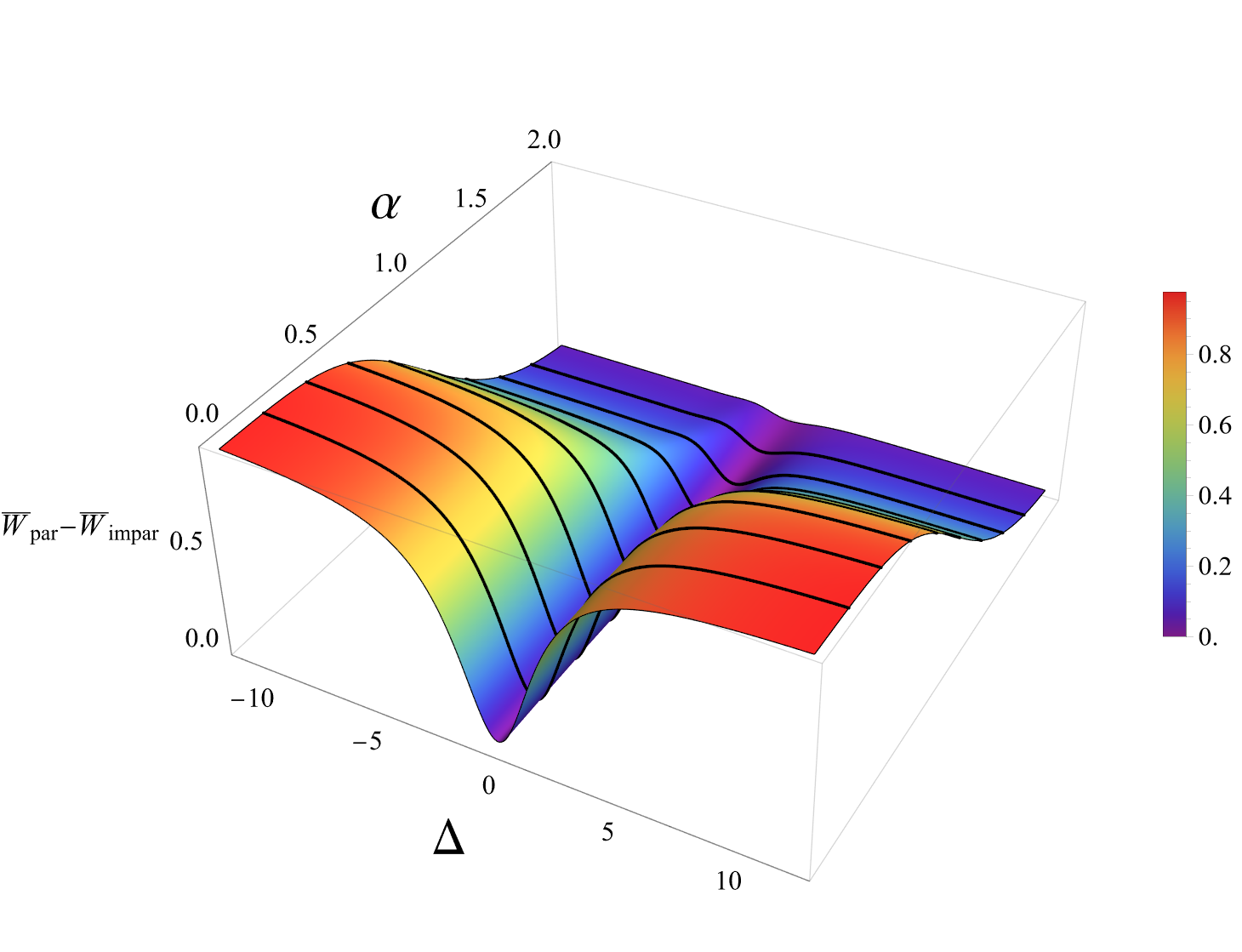}}
    \caption{Diferencia de las formas de línea cuando el estado inicial del átomo es el base, y el del campo son los estados gato de Schrödinger par e impar con $\alpha$ variando entre 0 y 2. Los valores de los parámetros son $\chi=0.5$ y $g=1.0$.}
    \label{fig7}
\end{figure}

Consideramos ahora que el estado inicial del campo es un estado gato de Schrödinger. Los estados gato de Schrödinger están definidos como
\begin{equation}\label{Superposition}
    \ket{\psi}  = \frac{1}{\mathcal{N}} \left( |\alpha \rangle + e^{\mathrm{i}\phi} |-\alpha \rangle \right),
\end{equation}
donde $\mathcal{N}$ es la constante de normalización dada por
\begin{equation} \label{Normalalization}
    \mathcal{N} = \sqrt{2 \left[1 +  e^{-2|\alpha |^2} \cos(\phi) \right]};
\end{equation}
es decir, son una superposición de dos estados coherentes con la misma amplitud, pero con diferentes fases\cite{Gerry_Book}; el nombre de estos estados fue acuñado por Schrödinger mismo en el año de 1935~\cite{Schrodinger_1935}. La distribución de probabilidad  fotones está dada por
\begin{align} \label{Distribution_Probability}
    P_n & = \left|\braket{n | \psi} \right|^2
    \nonumber                                 \\ &
    = \frac{2}{\mathcal{N}^2} \frac{e^{-|\alpha|^2}}{n!} |\alpha|^{2n}  \left[ 1 + (-1)^n  \cos(\phi) \right], \; n=0,1,2,\dots.
\end{align}
Nótese que cuando $\phi=0$, la probabilidad de que el número de fotones sea impar es cero ($P_{2n+1}=0, \; n=0,1,2,3,\dots$); mientras que para $\phi=\pi$, la probabilidad de que el número de fotones sea par es cero ($P_{2n}=0, \; n=0,1,2,3,\dots$). Por eso, con cierta frecuencia, a esos dos estados se les llama par e impar, respectivamente; así que los estados pares e impares son una superposición de dos estados coherentes con la misma amplitud, pero con fases opuestas~\cite{Gerry_Book}. En la figura~\ref{fig5}, presentamos la distribución de fotones cuando $\alpha=4$, para los estados par e impar.
Estudiamos ahora la inversión atómica promedio, las formas de línea, cuando el estado inicial del campo es un estado gato de Schrödinger, específicamente vamos a considerar los estados pares e impares. En la siguiente imagen, Fig.~\ref{fig6}, mostramos la diferencia en las formas de línea entre los estados gato de Schrödinger par e impar, variando $\alpha$ desde $0$ hasta $2$, cuando el átomo está inicialmente en el estado excitado; en esa gráfica $\chi=0.5$, valor que está dentro del rango de validez de la aproximación supuesta para que el Hamiltoniano \eqref{eq:1} sea valido, y hemos hecho $g=1.0$.
Observamos que para un número promedio de fotones $\bar{n}=|\alpha|^2<4$, es posible discernir entre un estado gato de Schrödinger par o impar, debido a la firma estadística del estado de vacío del campo electromagnético en $P_{2n}$. El efecto se pierde a medida que aumenta el número promedio de fotones, ya que la probabilidad de encontrar $n$ fotones en un estado coherente se concentra alrededor de $|\alpha|^2$.\\
Pasemos ahora a analizar qué sucede cuando el átomo está inicialmente en el estado base. En la Fig.~\ref{fig7}, mostramos la diferencia en las formas de línea de los estados gato par e impar, en el caso en que originalmente el átomo se halle en el estado base; nuevamente consideramos una variación de  $\alpha$ entre $0$ y $2$, y además, como en el caso anterior, $\chi=0.5$ y $g=1.0$.
Observamos que la diferencia en las formas de línea entre los estados pares e impares es más notoria que en el caso previo, debido a la existencia de una excitación adicional entre el estado base y el estado excitado, como se mencionó en la sección anterior. Sin embargo, esta diferencia en las formas de línea (cuando el átomo se encuentra en el estado base o excitado) no permite discernir entre estados pares e impares cuando el número promedio de fotones es $\bar{n} > 4$. La razón de esta imposibilidad es que el rango de validez del Hamiltoniano \eqref{eq:1} está limitado a valores de $\chi$ menores que $1$~\cite{Villanueva_2020}, y a que para numero promedio de fotones lo suficientemente grande, la distribución de probabilidad para los estados coherentes pares e impares son aproximadamente indistinguibles.

\section{Conclusiones}\label{Conclusión}
En este trabajo consideramos un átomo con un estado base, un primer estado excitado y estados superiores. El átomo interactúa con un campo electromagnético de un solo modo. Suponemos que el campo está aproximadamente en sintonía con la frecuencia de transición entre los dos primeros niveles del átomo, pero fuera de sintonía con los niveles cercanos.\\
Demostramos que las formas de línea, es decir, las formas de transición atómica medidas a través de la inversión atómica promedio $\overline{W}(\Delta)$ en función de la desintonía, permiten distinguir entre estados gato de Schrödinger pares e impares. Además, al inicializar el átomo en el estado base, se produce una excitación adicional entre el estado base y el estado excitado, lo que aumenta las diferencias entre las formas de transición atómica, y facilita dicha distinción. Sin embargo, el rango de validez del Hamiltoniano \eqref{eq:1} está limitada a valores de $\chi$ menores que $1$~\cite{Villanueva_2020}, lo que limita la capacidad de distinguir entre estados pares e impares cuando el número promedio de fotones es mayor que 4 ($\bar{n}=|\alpha|^2>4$).

\section*{Agradecimientos}
L. Hernández Sánchez agradece al Instituto Nacional de Astrofísica, Óptica y Electrónica (INAOE) y al Consejo Nacional de Ciencia y Tecnología (CONACyT) por la beca doctoral otorgada (No. CVU: 736710).
%
\end{document}